# An efficient pre-object collimator based on an x-ray lens

Erik Fredenberg* and Björn Cederström

*Department of Physics, Royal Institute of Technology,*

*AlbaNova, SE-106 91 Stockholm, Sweden*

Magnus Åslund

*Sectra Mamea AB, Smidesvägen 5, SE-171 41 Solna, Sweden*

Peter Nillius and Mats Danielsson

*Department of Physics, Royal Institute of Technology,*

*AlbaNova, SE-106 91 Stockholm, Sweden*

(Dated: December 9, 2008)

## Abstract

A multi-prism lens (MPL) is a refractive x-ray lens with one-dimensional focusing properties. If used as a pre-object collimator in a scanning system for medical x-ray imaging, it reduces the divergence of the radiation and improves on photon economy compared to a slit collimator. Potential advantages include shorter acquisition times, a reduced tube loading, or improved resolution. We present the first images acquired with an MPL in a prototype for a scanning mammography system. The lens showed a gain of flux of 1.32 compared to a slit collimator at equal resolution, or a gain in resolution of 1.31 − 1.44 at equal flux. We expect the gain of flux in a clinical set-up with an optimized MPL and a custom-made absorption filter to reach 1.67, or 1.45 − 1.54 gain in resolution.

## I. INTRODUCTION

Scanning systems for medical x-ray imaging have the advantage of efficient intrinsic scatter rejection.[1] Some of the available x-radiation is, however, blocked before the object by the pre-object collimator, which might lead to higher tube loadings or longer image acquisition times. Hence there is a need for more efficient use of the x-ray tube output.

X-ray optics have been suggested as a way to reach a better photon economy than is achievable with conventional methods. Capillary optics are similar to fibre optics for visible light, and bundles of capillaries can be placed as a grid after the object to reduce scattered radiation and improve on resolution.[2] It is a more efficient method for scatter rejection than conventional grids or air gaps, but the size of the field is limited due to manufacturing constraints, and absorption of primary quanta is still present. Another option is to increase the flux by using capillaries to gather radiation before the object.[3] In that case, divergence of the radiation at the exit side of the capillaries distorts the resolution so that a combination with a diffracting crystal is necessary; the capillaries increase the flux to the crystal and the limited acceptance angle of the crystal reduces the divergence of the beam.

The multi-prism lens (MPL) is a refractive x-ray lens. It distinguishes itself from other x-ray lenses, such as parabolic lenses, compound refractive lenses, and fresnel lenses, by a simplified manufacturing process, and a tunable focal length.[4] Measurements have been performed at synchrotron facilities for various lens materials such as epoxy, silicon, and lithium.[5,6] Studies have also been performed with x-ray tubes using lenses made of e.g. vinyl, plexiglas, and epoxy.[4,7,8]

There are at least two different ways an MPL can be employed in scanning systems for medical x-ray imaging; (1) as an energy filter to optimize the x-ray spectrum, and (2) as a focusing pre-object collimator that reduces the divergence of the beam and increases the utilization of the available x-rays compared to a slit collimator. Previous investigations of the first aspect have shown that the MPL might be favorable compared to absorption filtering techniques.[8,9] In this work, we investigate the latter aspect of the lens. The MPL can be expected to have no negative effect on the resolution in the non-focusing direction, which would be the case for capillary optics, and it is more convenient from a manufacturing point of view than other types of refractive or diffractive x-ray optics. Furthermore, the one-dimensional focus of the MPL matches the linear detectors found in some scanning

systems.[1,10,11]

We present the first images acquired with an MPL in the beam path. Flux and resolution of a model mammography set-up have been compared to a set-up with slit collimation to investigate the gain. Ray-tracing has been used for modeling and predictions. Potential challenges for implementation in a clinical system have been considered.

## II. MATERIALS AND METHODS

### A. The MPL collimator

An MPL consists of two rows of prisms (lens-halves) facing each other at an angle and symmetrically arranged around the optical axis (Fig. 1, left).[4] Peripheral rays encounter more prisms on the way through the lens than central ones, and therefore experience a greater refraction — a focusing effect is achieved in one dimension. Since MPL's consist of only flat surfaces, manufacturing is relatively easy, and the lenses are therefore suitable for small-scale applications.

According to previous reports,[4] the focal length of an MPL is calculated as

$$F = \frac{d_\text{g} \tan\theta}{2N\delta(E)},  \quad (1)$$

where $d_\text{g}$ is the offset from the optical axis of one lens-half at the exit side of the lens, $\theta$ is the prism base angle, $N$ is the number of teeth, and $\delta$ is the decrement from unity of the real part of the index of refraction. The focal length can be tuned by changing $d_\text{g}$ in an otherwise fixed geometry.

For this study, we have investigated a slightly modified version of the MPL. The angle between the lens-halves is small, but instead the prism height decreases towards the exit side of the lens (Fig. 1, right). Additional small gaps ($d_\text{s1}$ and $d_\text{s2}$) were introduced between the lens-halves at the entrance and exit sides. These gaps can be used to trim the collimator width, and also the focal length because $d_\text{g} = d_\text{t} + (d_\text{s2} - d_\text{s1})/2$, where $d_\text{t}$ is the tooth height. Note that configurations other than $d_\text{s2} = d_\text{s1}$ affects the aperture, and $d_\text{g}$ is restricted within a small interval. Consequently, the modified lens is only slightly tunable, but is instead short. The decrement from the optical axis of each tooth is $d_\text{i} \approx d_\text{t}/N$, and the length is

$$L = \sum_{n=0}^{N-1} \frac{2d_\text{t}}{\tan\theta}\left(1 - n\frac{d_\text{i}}{d_\text{t}}\right) = \frac{d_\text{t}}{\tan\theta}(N+1), \quad (2)$$

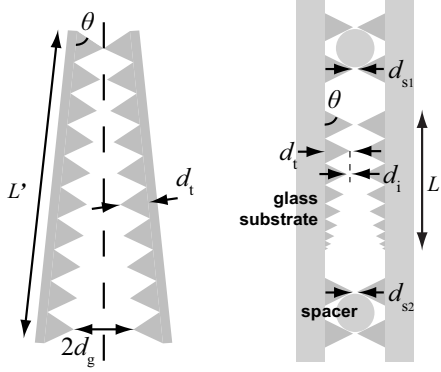

FIG. 1: A schematic of the original MPL (left), and the modified lens evaluated in this study (right). The lens has the following design parameters: $d_t = d_g = 70$ μm, $d_i = 0.12$ μm, $d_{s1} = d_{s2} = 10$ μm, $\theta = 54.7°$, $F = 108$ mm at $E = 18$ keV, and $N = 606$ teeth. The length of the modified lens is $L = 30$ mm, which is approximately half that of the original lens, $L'$. Each lens-half is mounted on a glass substrate. Spacers at the entrance and exit sides keep the lens-halves an appropriate distance apart.

thus approximately half the length of a standard MPL, which is close to $2d_t N/\tan\theta$.[4] Henceforth, the modified MPL will be referred to as the MPL.

We have studied an MPL set-up with source-to-lens and source-to-image distances of 507 and 640 mm. A thin lens with a focal length of $F = 108$ mm in such a set-up images the source at the detector plane, according to the Gaussian lens formula. We have chosen an epoxy lens with this focal length at 18 keV, as it is close to the mean energy of typical mammography spectra. This is a simple optimization scheme, and it is possible that the performance can be improved slightly with a better optimized focal length. Since $\delta \propto E^{-2}$ in the x-ray region, $F \propto E^2$ according to Eq. (1). Lower-energy photons therefore have beam waists between the lens and detector, whereas higher-energy photons are focused behind the detector.

A tooth height of $d_t = 70$ μm, and a prism base angle of $\theta = 54.7°$ were chosen for the lens. According to Eq. (1), the required number of teeth for $F = 108$ mm at 18 keV and $d_{s2} = d_{s1}$ is then $N = 606$. The decrement from the optical axis of each tooth becomes $d_i = 0.12$ μm and the lens length $L = 30$ mm.

The MPL used in the experimental part of this study was manufactured by commercial companies as part of the HighReX EU-project.[12] Anisotropic etching was used to produce a lens master from a single-crystalline silicon wafer so that the $<111>$ lattice planes define the prism sides (Silex Microsystems, Järfälla, Sweden). The etched silicon masters only resolved teeth down to 10 $\mu$m height, so the number of usable teeth was limited to 519. The masters were replicated with UV-embossing in epoxy, and the epoxy lens-halves were mounted on 500 $\mu$m thick pyrex glass substrates (Heptagon, Zürich, Switzerland). Finally, the lens was assembled by gluing the halves on quartz optical fibres, which worked as spacers and absorption filters on the entrance and exit sides of the lens (Fig. 1). The fibre diameter was 125 $\mu$m so that $d_{s2} \approx d_{s1} \approx 10$ $\mu$m, but the gaps were associated with some uncertainty because of the gluing process.

### B.  Gain compared to a slit collimator

This study compared an experimental model of a scanning digital mammography set-up with a single pre-breast slit collimator (the slit set-up) to a set-up with an MPL collimator (the MPL set-up). We used measurements and ray-tracing simulations with methods described in the coming two sections. Figure 2 outlines the two set-ups, both with source-to-image distances of 640 mm. A source-to-slit distance of 535 mm was chosen so that it approximately matches the exit side of the lens in the MPL set-up with a source-to-lens distance of 507 mm. A full mammography system would employ several collimator slits in a multi-slit geometry, but the slits can be exchanged for an array of MPL's so the extension from our model set-ups to full systems is straightforward. One example of a clinical mammography system based on a scanning multi-slit geometry is the Sectra MicroDose Mammography (MDM) system (Sectra Mamea AB, Solna, Sweden).[1,10,11] The MDM system is similar, but not identical to the slit set-up.

For all comparisons, the energy spectra of the two set-ups need to be approximately equal so that the dose efficiencies are similar. Therefore, we matched the spectrum hardness in terms of the half-value layer (HVL) by applying appropriate absorption filtering to both set-ups with methods described in the coming two sections.

It can be apprehended that scattering in the MPL might degrade the low-contrast performance of the set-up. We do, however, expect most scattered photons to be absorbed in the lens because of the wide angular spectrum of Compton scattering. Moreover, previous studies indicate that scanned systems are relatively insensitive to scatter,[1] which means that any additional scattering should be rejected by the geometry of the MPL set-up and not influence the performance. Nevertheless, it is important to verify these hypotheses, and to that end we measured and compared the contrast of a low-contrast object in the two set-ups.

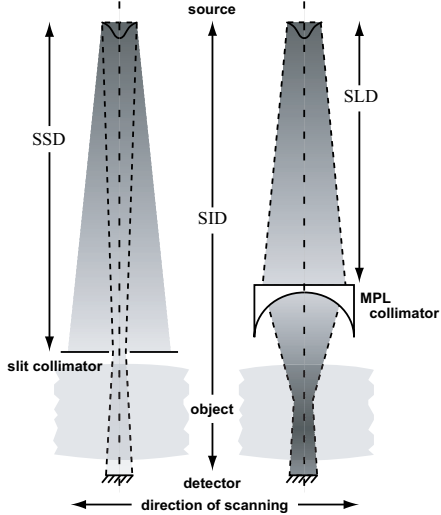

FIG. 2: Schematics of the slit (left) and MPL (right) set-ups compared in this study. The source-to-image distance refers to the detector plane and was SID = 640 mm. The source-to-slit and source-to-lens distances were SSD = 535 and SLD = 507 mm.

TABLE I: Definitions of the resolution and gain metrics that were used to compare the MPL and slit set-ups.

| | |
|---|---|
| $MTF_{0.5}$, $MTF_{0.1}$ | The spatial frequencies at MTF = 0.5 and MTF = 0.1 in the center of a 45 mm breast. |
| $G_f$ | Gain of flux. |
| $G_{0.5}$, $G_{0.1}$ | Gain in $MTF_{0.5}$ and $MTF_{0.1}$ frequencies. |

The MPL set-up can be designed either to increase the flux or the spatial resolution compared to the slit set-up. To quantify the improvement in photon economy, we used the gain of flux ($G_f$), which is the ratio of the intensities in the detector plane of the two set-ups. $MTF_{0.5}$ and $MTF_{0.1}$ refer to the spatial frequencies when the modulation transfer function (MTF) has dropped to 0.5 or 0.1. The gain in resolution ($G_{0.5}$ or $G_{0.1}$) is defined as the ratio between $MTF_{0.5}$ or $MTF_{0.1}$ of the two set-ups. These definitions are summarized in Table I.

The comparison between the set-ups was conducted by evaluating $G_\text{f}$ at a matching resolution ($G_{0.5} \approx 1$), and $G_{0.5}$ and $G_{0.1}$ at matching flux ($G_\text{f} \approx 1$) using measurements and ray-tracing simulations at a distance from the detector corresponding to the center of a 45 mm breast. The gap between the lens-halves was varied in the ray-tracing model to investigate the effect on these metrics. It can be expected that the resolution in the MPL set-up varies with depth in the breast differently than for a slit collimator, and the resolution in terms of $\text{MTF}_{0.5}$ and $G_{0.5}$ as a function of distance to the detector was therefore simulated with ray-tracing. Finally, we expect misalignment of the lens in a clinical set-up of up to 0.1° to the optical axis, either at the time of manufacturing or during use. The change in gain was therefore simulated for a range of angular displacements by changing the lateral displacement of the source from the optical axis.

### C. Experimental set-up and measurements

The experimental set-up that was used to represent the MPL and slit set-ups, is shown in Fig. 3. In the MPL configuration, the lens was mounted on precision stages for lateral translation and tilting with the entrance side 507 mm from the source. If instead used as a slit set-up, the lens was exchanged for a 2 mm thick tantalum collimator slit 535 mm from the source, corresponding to the exit side of the lens. The width of the slit determined the resolution in the direction of scanning, and it was adjusted to match either the flux or the resolution of the MPL set-up.

We used a tungsten target x-ray tube (Philips PW2274/20), which is water cooled and can be run continuously. The acceleration voltage for all measurements was 33 kV, and the anode angle was 2.2°. We measured the spot size in the scan direction of the set-ups with an edge scan to be 425 $\mu$m full width at half maximum (FWHM) and approximately trapezoidal with a base-to-FWHM ratio of 1.3. The spot size perpendicular to the scanning was specified to 400 $\mu$m by the manufacturer, but it is of less importance since the set-ups should be equal in that direction. Both set-ups were filtered with 3 mm of polymethyl methacrylate (PMMA) to account for the compression plate in a realistic mammography system. Additional aluminum filtration was added to the slit set-up to match the HVL of the unfiltered MPL.

To acquire images, we used an edge-on silicon-strip detector with a 50 $\mu$m strip-pitch,

wire-bonded to a 128-channel pulse-counting application specific integrated circuit (ASIC). All pulses below a certain detection threshold are rejected by the ASIC, and the detector assembly is thus photon-counting with virtually no electronic noise present. Detailed descriptions of the detector have been published previously.[10,11] The detector strip-pitch and the x-ray tube focal spot size determined the resolution perpendicular to the direction of scanning for both set-ups. The x-ray tube current was 55 mA when images were acquired.

To measure energy spectra, we replaced the silicon-strip detector with a CZT compound solid-state detector (Amptek XR-100T-CZT) at the same distance from the source. It has a near 100% detection efficiency, negligible hole tailing, and an energy resolution better than 0.5 keV in the considered energy interval. The detector was collimated and the tube current reduced to 20 mA to reach a count rate of $\sim$ 2000 s$^{-1}$ without pile-up. A 60 s acquisition time was regarded sufficient with these settings. The air kerma was calculated from the energy spectrum, and the HVL was found as the aluminum thickness that reduces the kerma by a factor of two.

In front of the detector, a motorized stage with a precision of 10 $\mu$m was used to scan a knife edge to obtain the MTF 618 mm from the source, corresponding to a plane in the middle of a 45 mm breast. The point spread function (PSF) was found as the derivative of the edge scan profile and the MTF as the fast Fourier transform of the PSF.

Alternatively, an object could be placed on the stage, which was then scanned across the beam to acquire an image. A bar-pattern phantom (CIRS Inc., Norfolk, Virginia) was used to visualize the resolution limits of the set-ups. To measure low-contrast performance, we used images of a 3.17 mm diameter artificial tumor in a 45 mm thick MEDI-TEST phantom (Domaine Technologique de Saclay - Immeuble Azur, Saclay, France) made of tissue equivalent material corresponding to 50% glandularity. The average glandular dose was calculated by applying normalized glandular dose coefficients[13] to the measured spectrum. As a reference, the contrast of the tumor was measured with a Sectra MDM system to 1.21% at 32 kV.

### D. Ray-tracing simulation and modeling

We simulated the set-up in a custom-made ray-tracing model using the MATLAB software package (The MathWorks Inc., Natick, Massachusetts). 2 000 rays per keV in steps of 1 keV from 5 to 35 keV were traced through the MPL. Origins of the rays were randomly distributed over the area of the source with probabilities corresponding to the measured source shape. The lens epoxy was assumed to be $C_{18}H_{20}O_3$ with density $\rho = 1.1$ g/cm$^3$, and the quartz optical fibres were $SiO_2$ with $\rho = 2.2$ g/cm$^3$. Air and the pyrex substrates were modeled with published compositions.[14] The refractive index was found from the real part of the atomic scattering factors ($f_1$),[15] and out-of-range energy values were extrapolated with a first degree polynomial in $f_1$. Absorption was calculated with published elemental absorption coefficients.[14] We implemented the limit in number of usable teeth in the model, but other possible lens imperfections such as surface roughness, air bubbles in the epoxy, and impurities and uncertainties in the lens materials were excluded. Scattering in the lens was treated as absorption.

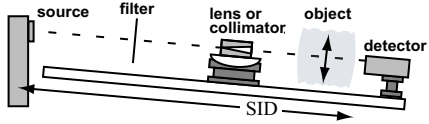

FIG. 3: The experimental set-up where the lens could be exchanged for a slit. We used a tungsten target x-ray tube with a 425 $\mu$m spot size in the scanning (vertical) direction. The detector is either a 128-channel silicon-strip detector or a CZT spectrometer. Absorption filters could be optionally inserted in the beam path.

The strip detector was modeled with pure silicon and an effective thickness of 7.1 mm. All charge released by photo-electric events was detected, and the detection threshold could be varied. Scattering in the detector was assumed not to contribute to the detected signal, but only to filtration of the beam through the detector. The tantalum slit was taken to have infinite absorption.

The ray-tracing model was verified with experimental data using the shape and energy spectrum of the experimental source as input. We expect the largest uncertainties in the lens construction to be on $d_{s1}$ and $d_{s2}$, and these were adjusted in the model to find close agreement with experimental results in terms of HVL, resolution, and gain. The slit widths in the experiment were also associated with some uncertainty, and the modeled slit widths were therefore chosen to match the measured resolution.

The verified model was used to predict the MPL performance in a clinically realistic set-up with $d_{s1}$ and $d_{s2}$ adjusted for best possible performance with the current lens. A published spectrum from a 33 kV mammography tungsten target x-ray tube[16] was used as input since the steep anode-angle of the experimental set-up leads to substantial self-filtration and a correspondingly harder spectrum than is common in mammography. The source shape was equal to the experimental source.

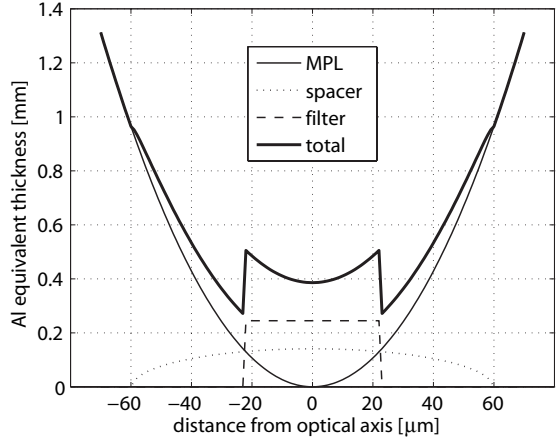

FIG. 4: The projected amount of material along the optical axis for a $d_{s2} = d_{s1} = 0$ $\mu$m MPL collimator. The lens, the spacers, and a 245 $\mu$m thick aluminum filter are shown together with the summed projection for all components in the assembly. The profiles are normalized to aluminum equivalent thickness in terms of HVL for a clinical mammography spectrum.

In this clinically more realistic case, the slit set-up was filtered with 0.50 mm aluminum, and the MPL set-up was adapted to match the resulting HVL. This particular filtering was chosen since it corresponds closely to what is used clinically in the Sectra MDM system.[10] Figure 4 shows the projected amount of material along the optical axis for the different components in a $d_{s1} = d_{s2} = 0$ $\mu$m MPL collimator, normalized to aluminum equivalent thickness in terms of HVL for the clinical spectrum. The lens itself has an approximately parabolic projection, and the transmission is therefore higher in the central parts of the lens than in the periphery, although the circular profile of the spacers levels out the non-uniformity somewhat. A 245 $\mu$m thick aluminum filter was therefore placed in the center of the lens aperture, and the width of the filter was adapted according to the lens gap to reach the HVL of the slit set-up.

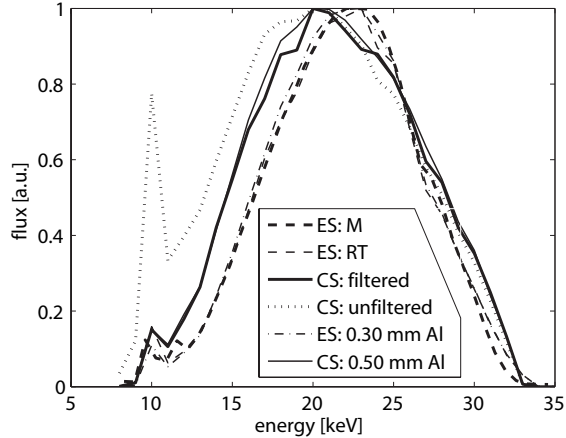

FIG. 5: The energy spectra of the MPL set-ups in the detector plane; measured (M) and ray-traced (RT) in the experimental set-up (ES) for a $(d_{s1}, d_{s2}) = (30, 20)$ μm MPL collimator. Ray-tracing results for a clinical set-up (CS) using a mammography spectrum as input to a $(d_{s1}, d_{s2}) = (0, 0)$ μm MPL are presented with and without a $245 \times 44$ μm aluminum filter. Also shown are the experimental and clinical spectra filtered with 0.30 and 0.50 mm aluminum.

## III. RESULTS

### A. Validation of the experimental set-up

The transmitted energy spectrum of the MPL set-up is shown in Fig. 5 together with the experimental spectrum filtered with 0.30 mm aluminum. The aluminum filtered spectrum matched the MPL spectrum within 5% in terms of HVL and should have approximately the same dose efficiency.

A 140 μm collimator slit was found to match the flux of the MPL set-up, and Figure 6 shows images of the 3.17 mm diameter tumor in the MEDI-TEST phantom acquired with the two set-ups. Relatively high exposure parameters of 1.7 mGy and 27 mAs were used to reduce the influence of statistical noise. Visually, the two images are similar in quality, and have equal non-uniformities. The mean pixel counts in the images is matched within 1.2%, the difference in signal-to-noise ratio within 2.2%, and the dose within 2.0%. We measured the contrast between the tumor and background to 1.09% and 1.12% in the MPL and slit set-ups, thus a difference of 3%. Compared to the reference measurement with the MDM system, the contrast of the experimental set-ups is less than 10% lower.

TABLE II: Measured (M) results are presented in the top row. Shown in the middle row are ray-tracing (RT) results from the experimental set-up (ES) with gaps ($d_{s1}$ and $d_{s2}$) adjusted to provide a close match to the experimental results. The bottom row illustrates a best-case scenario in terms of gain in a clinical set-up (CS) with a mammography spectrum as input, a custom-made absorption filter, and no gap between the lens-halves. The gain was calculated at matching resolution and matching flux, meaning that $G_{0.5} \approx 1$ and $G_f \approx 1$ respectively. Cf. Table I for a summary of the metrics.

|  | set-up | lens $(d_{s1}, d_{s1})$ | filter [$\mu$m] | HVL [mm] | MTF$_{0.5}$ [mm$^{-1}$] | MTF$_{0.1}$ [mm$^{-1}$] | matching resolution $G_f$ | $G_{0.5}$ | $G_{0.1}$ | matching flux $G_f$ | $G_{0.5}$ | $G_{0.1}$ |
|---|---|---|---|---|---|---|---|---|---|---|---|---|
| M: | ES | experimental | – | 0.57 | 4.6 | 7.9 | 1.32 | 1.00 | 1.08 | 0.98 | 1.31 | 1.44 |
| RT: | ES | (30, 20) | – | 0.57 | 4.8 | 7.9 | 1.26 | 1.06 | 1.09 | 0.93 | 1.35 | 1.43 |
| RT: | CS | (0, 0) | 44 × 245 | 0.51 | 5.8 | 9.7 | 1.67 | 1.00 | 1.01 | 1.00 | 1.45 | 1.54 |

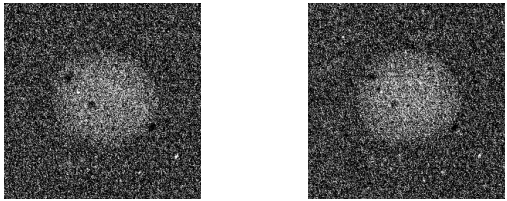

FIG. 6: Images of a 3.17 mm diameter low-contrast tumor in the MEDI-TEST phantom at approximately equal exposure, HVL, and dose; the MPL set-up (left), and the slit set-up (right).

### B. Gain and resolution measurements

Figure 7 shows the MTF in the focusing direction measured 618 mm from the source, which corresponds to the center of a 45 mm breast. The measured MTF$_{0.1}$ and MTF$_{0.5}$ frequencies are presented in the first row of Table II. The MTF of the MPL was matched by a 110 $\mu$m slit so that $G_{0.5} = 1$, but the resolution was better at higher frequencies, resulting in a $G_{0.1}$ larger than unity. $G_f$ was measured to be 1.32. The MTF of the 140 $\mu$m collimator slit that matched the flux of the MPL set-up is also shown in Fig. 7. In this case, $G_{0.5}$ and $G_{0.1}$ were measured to be 1.31 and 1.44 (cf. Table II).

The higher resolution of the MPL set-up is further illustrated in the top part of Fig. 8, showing images of the bar-pattern phantom for both set-ups that were acquired with 22 mAs per pixel. The MPL set-up was able to correctly reproduce up to 10 lp/mm, whereas the slit set-up exhibited aliasing above 7 lp/mm. The lower part of Fig. 8 shows the bar-pattern phantom in the non-focusing direction, perpendicular to the scanning. No difference was seen between the set-ups, which shows that the MPL does not degrade the limiting resolution in this direction.

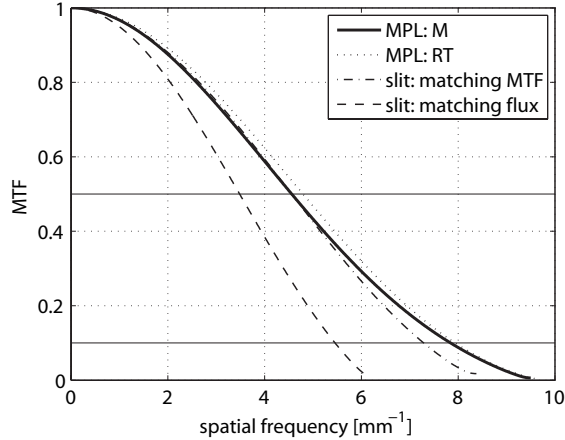

FIG. 7: The measured (M) MTF of the MPL set-up in the focusing direction compared to 110 and 140 $\mu$m collimator slits matching the MTF and flux of the set-up. Also shown is the MTF predicted by ray-tracing (RT) for a $(d_{s1}, d_{s2}) = (30, 20)$ $\mu$m MPL collimator. The horizontal lines indicate $\text{MTF}_{0.5}$ and $\text{MTF}_{0.1}$.

### C. Comparison of ray-tracing to measurements

A $(d_{s1}, d_{s2}) = (30, 20)$ $\mu$m MPL set-up in the ray-tracing model was found to predict the measured energy spectrum well with no difference in HVL. As can be seen in Fig. 7, the resolution is also closely matched with an $\text{MTF}_{0.5}$ frequency 0.2 mm$^{-1}$ higher than measured according to Table II, row 2.

The measured resolution of the 110 and 140 $\mu$m slits was matched in the ray-racing model by 102 and 137 $\mu$m slits, and the gain of the MPL collimator in relation to these is shown in Table II, row 2. Compared to the measured gain in row 1, $G_f$ is 5% less in the ray-tracing (1.26 compared to 1.32), whereas $G_{0.5}$ is 3% greater (1.35 compared to 1.31).

Figure 9 is divided into two parts, of which the first one visualizes the experimental set-up, i.e. the relation between the MPL measurement and the ray-tracing result for the

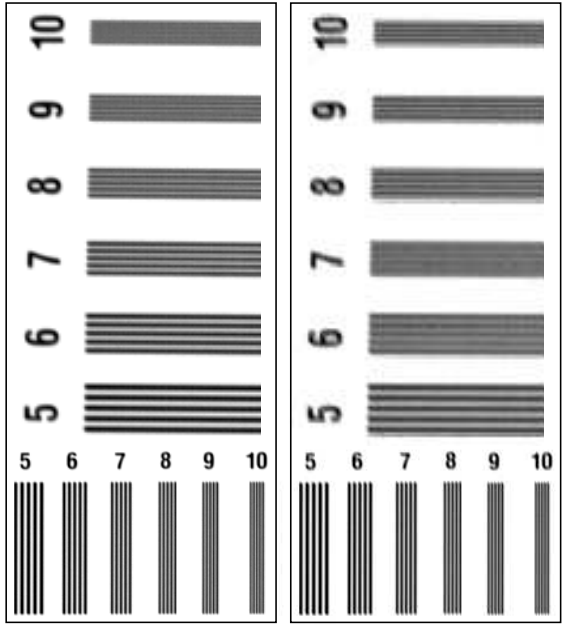

FIG. 8: Images of the bar-pattern phantom at approximately equal flux, exposure, and HVL; the MPL set-up (left), and the slit set-up (right). At the top are images acquired in the focusing direction, compared to the non-focusing direction at the bottom. The numbers indicate lp/mm.

$(d_{s1}, d_{s2}) = (30, 20)$ $\mu$m lens that best fitted the measurement. The figure shows MTF$_{0.5}$ as a function of flux, normalized to the matching-resolution slit on the slit-collimator line. This slit was 102 $\mu$m in the ray-tracing model, but measured as 110 $\mu$m (shown with parenthesis in the figure). The MPL- and slit-collimator points are all indicated with a square or a circle for measurement and ray-tracing respectively. $G_f$ or $G_{0.5}$ can be found as the ratio of the MPL-collimator point to points with equal resolution or equal flux on the slit-collimator line. If there were perfect agreement between measurement and ray-tracing, corresponding squares and circles would coincide, and the equal resolution and equal flux slit-collimator points would be located horizontally to the left and vertically below the MPL-collimator points. Hence, the distance between the MPL-collimator points shows the discrepancy in gain between model and measurement, and the displacement of the slit-collimator points indicate that $G_f$ and $G_{0.5}$ were not exactly unity when matching the slits.

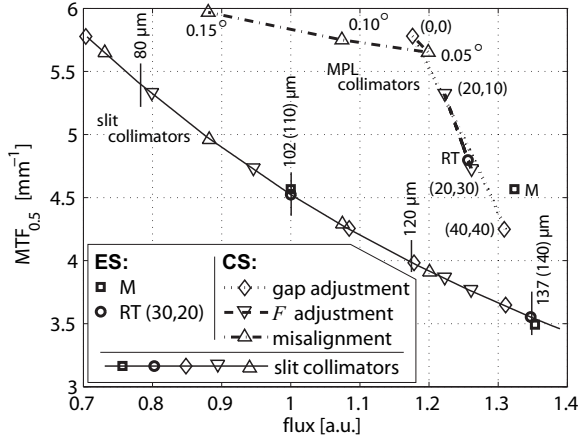

FIG. 9: Resolution (MTF$_{0.5}$) versus flux, normalized to the equal-resolution slit (102 and 110 $\mu$m for ray-tracing and measurements respectively). Firstly, this figure visualizes the experimental set-up (ES) by measurement (M) and ray-tracing (RT) results. The MPL-collimator points have corresponding equal flux and equal resolution points indicated with an identical symbol on the slit-collimator line. A few slit widths used in the ray-tracing model are marked out on the line, with corresponding measured widths in parenthesis if applicable. The gain presented in Table II is the ratio of an MPL-collimator point to a slit-collimator point. Secondly, three different effects calculated with ray-tracing in the clinical set-up (CS) are shown; (1) diamonds connected by a dotted line indicate change in gap ($d_{s2} = d_{s1}$), (2) standing triangles connected by a dashed line are $d_{s2}$ adjusted for a change in focal length ($F$), and (3) lying triangles connected by a dashed-dotted line show increasing angular displacement. The numbers in parenthesis represent lens gaps as ($d_{s1}, d_{s2}$), and the angles indicate the misalignment from the optical axis.

### D. Ray-tracing predictions in a clinical set-up

The second part of Fig. 9 concerns ray-tracing results with the clinical spectrum as input. Three different effects are shown in the figure, the first two being results for a range of lens gaps ($d_{s1}, d_{s2}$). Firstly, a change in gap with $d_{s2} = d_{s1}$, corresponding to the width of a slit collimator, is illustrated with a dotted line labeled "gap adjustment." Secondly, a dashed line labeled "$F$ adjustment" indicates the change that occurs if $d_{s2} = d_{s1} \pm 10$ $\mu$m, which corresponds to a change in focal length of $\pm 7\%$. In each point, the diameter of the spacer was adapted to the gap, and the width of the filter was adjusted to reach an HVL of 0.51 mm, corresponding to 0.50 mm aluminum filtration in the slit set-up. The lines were calculated as a least square fit to several points, but only the end points are shown as diamonds and standing triangles respectively. Matching flux and resolution points are indicated on the slit-collimator line with equal symbols.

Figure 9 shows that the highest gain in both flux and resolution is found for a $(d_{s2}, d_{s1}) = (0,0)$ μm spacing. Spectra calculated for that lens are presented in Fig. 5, with and without a $44 \times 245$ μm aluminum filter, and the effect of the filter is evident. A 73 μm slit matched the resolution of the filtered MPL, with a gain of flux of $G_f = 1.67$, whereas a 120 μm slit matched the flux, with gains in resolution of $G_{0.5} = 1.45$ and $G_{0.1} = 1.54$. The results are presented in the last row of Table II.

The third effect illustrated in Fig. 9 is the sensitivity to misalignment in terms of resolution and transmission of the $(d_{s1}, d_{s2}) = (0,0)$ μm MPL. It is shown as lying triangles, connected by a dashed-dotted line labeled "misalignment" The filter width was 44 μm for all points. $MTF_{0.5}$ was almost constant for angular displacements within the alignment tolerance of $0.1°$ and increased approximately 1%. The loss of transmitted flux was approximately 10% in the same regime.

Figure 10 shows $MTF_{0.5}$ as a function of distance from the detector plane for the $(d_{s1}, d_{s2}) = (0,0)$ μm MPL collimator compared to the 120 μm matching flux slit. A maximum in the resolution was seen for the MPL approximately 50 mm from the detector plane, whereas the slit resolution was monotonically decreasing towards the detector. These two curves yield $G_{0.5}$ by division, which peaks 30 mm from the detector plane as can be seen in the figure.

## IV. DISCUSSION

### A. Measurements and ray-tracing comparisons

The slit could only be adjusted in steps of 10 μm, which resulted in a $G_f$ that differed 2% from unity when matching the flux of the experimental MPL and slit set-ups. There were also differences in HVL, tumor contrast, and signal-to-noise ratio between the two set-ups of 5%, 3%, and 2% respectively. We expect these variations to be negligible for our purposes and the set-ups should be comparable. The slightly lower tumor contrast compared to the reference measurement can be explained by the harder spectrum and the absence of a post collimator in the experimental set-ups.

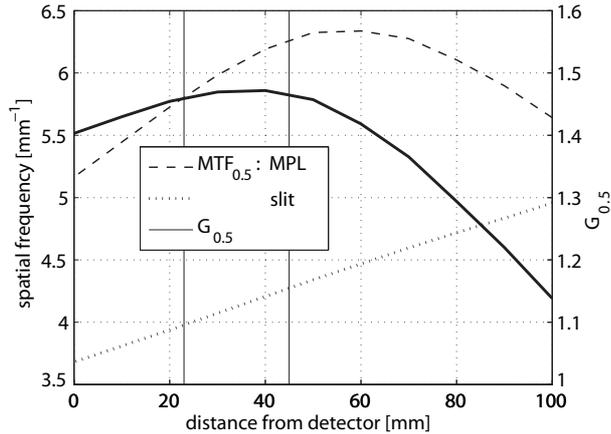

FIG. 10: Resolution ($MTF_{0.5}$) as a function of distance from the detector plane for an MPL collimator and a slit at matching flux. Also shown is the gain in resolution ($G_{0.5}$). The vertical lines correspond to the center and top of a 45 mm breast.

The ray-tracing model indicated that the slits used for the measurement were somewhat smaller than expected. Such a deviation is not unlikely and can be explained by uncertainties in the widths of the spacers used to attain the slit opening, or a slight tilt of the slit in relation to the optical axis. $G_{0.5}$ (and $MTF_{0.5}$) was predicted by ray-tracing to be slightly higher than was actually measured, whereas $G_f$ was lower. Note that the ray-tracing model tends to underestimate the performance of the lens as the difference in $G_f$ is larger than in $G_{0.5}$.

Although fairly small, the gain discrepancies should be kept in mind. It is likely that they are owing mainly to deviations in the exact epoxy composition, as it was not available, and an approximate formulation had to be used. A higher epoxy transmission would yield a higher $G_f$ and a lower resolution due to more transmission at the lens periphery. Rayleigh scattering and Mie scattering on structural imperfections in the lens such as air bubbles would also yield a higher transmission since scattered photons miss teeth, and a lower resolution due to a less convergent beam. Compton scattered photons at large angles can not be expected to contribute much to the deviation since they are likely to miss the detector.

### B. Ray-tracing predictions

It is clear from Fig. 9 that it is possible to increase the flux at the cost of a lower resolution by increasing $d_{s1}$ and $d_{s2}$. The gain in both flux and resolution decreases, however, when moving away from zero, and for large gaps the lens corresponds to a slit with unity gain. The difference between $d_{s1}$ and $d_{s2}$ corresponds to a focal length adjustment, but as can be seen in the figure, an adjustment within the considered interval has almost no effect.

The aluminum filter in the center of the lens provided fairly efficient filtering, but as can be seen in Fig. 4, the collimator profile is still not uniform. We expect that much is to gain by a better optimized filter with a decreasing thickness towards the periphery.

For a small misalignment of the MPL, the same effect was seen as when increasing the gaps; a higher flux and a lower resolution. For larger angular displacements, the resolution increased and the transmission decreased due to more vignetting (absorption in the lens sides), and a correspondingly narrower transmitted beam. Although the decrease in transmission is manageable for an individual lens, a potential problem arises if the MPL's in an array are misaligned with each other. Since the transmission curve is not linear, the change in transmission might vary among the lenses if the array is displaced. In particular, if two lenses are aligned with respectively positive and negative angles to the optical axis, the transmission for the lenses will change in opposite directions.

Figure 10 shows that within a 45 mm breast, the resolution of the MPL set-up varied with depth similar to the slit collimator so that the gain in resolution was approximately constant. Due to the resolution maximum, the gain decreased at larger distances and came close to unity in the plane of the slit. The MPL set-up is thus most efficient for small and average breast thicknesses.

## V. CONCLUSIONS

We have investigated a refractive multi-prism x-ray lens (MPL) for improved photon economy in medical x-ray imaging, in particular mammography. The MPL is a promising technology to reduce the acquisition time or improve on resolution.

We compared two scanning mammography systems with respectively an MPL and a slit as pre-object collimator. The HVL and low-contrast performance of the systems were found to be similar. We measured the gain of flux of the MPL to 1.32 at a better or equal MTF as the slit set-up, and the gain in resolution to 1.31 – 1.44 at a similar flux. The gain in resolution was further illustrated by images of a bar-pattern phantom, where at least three more lp/mm where seen in the MPL images.

A ray-tracing model was presented to predict results by the MPL set-up. The gaps between the lens-halves in the experimental MPL were considered uncertain and were therefore adjusted in the model to find a close match to the measurements. The remaining deviations in resolution and gain are within 5%, and we expect these to be covered by lens imperfections and model approximations.

Using the ray-tracing model, we estimated that an optimal lens with a proper absorption filter would yield a gain of flux of 1.67, or a gain in resolution of 1.45 – 1.54 in a clinical set-up.

A misaligned lens improved the MTF slightly, but reduced the transmission approximately 10% at a 0.1° displacement. This reduction is not severe but needs to be taken into account for clinical implementations. The model also showed that the resolution of the MPL set-up decreases close to the exit side of the lens, and it is therefore most efficient for smaller and average sized breasts.

**Acknowledgments**

We gratefully acknowledge Alexander Chuntonov and Mats Lundqvist for setting up the silicon-strip detector, and Staffan Karlsson for coordinating lens fabrication. This research was funded in part by the Swedish Research Council.

---

[*] E-mail: `fberg@mi.physics.kth.se`

[1] M. Åslund, B. Cederström, M. Lundqvist, and M. Danielsson. Scatter rejection in multi-slit digital mammography. *Med. Phys.*, 33:933–940, 2006.

[2] C.D. Bradford, W.W. Peppler, and R.E. Ross. Multitapered x-ray capillary optics for mammography. *Med. Phys.*, 29(6):1097–1108, 2002.

[3] F.R. Sugiro, D. Li, and C.A. MacDonald. Beam collimation with polycapillary x-ray optics for high contrast high resolution monochromatic imaging. *Med. Phys.*, 31(12):3288–3297, 2004.

[4] B. Cederström, R. Cahn, M. Danielsson, M. Lundqvist, and D. Nygren. Focusing hard x-rays with old LP's. *Nature*, 404:951, 2000.

[5] B. Cederström, M. Lundqvist, and C. Ribbing. Multi-prism x-ray lens. *Appl. Phys. Lett.*, 81(8):1399–1401, 2002.

[6] D.A. Arms, E.M. Dufresne, R. Clarke, S.B. Dierker, N.R. Pereira, and D. Foster. Refractive optics using lithium metal. *Rev. Sci. Instrum.*, 73(3):1492–1494, 2002.

[7] W. Jark. A simple monochromator based on an alligator lens. *X-Ray Spectrom.*, 33:455–461, 2004.

[8] E. Fredenberg, B. Cederström, M. Åslund, C. Ribbing, and M. Danielsson. A tunable energy filter for medical x-ray imaging. *X-Ray Optics and Instrumentation*, 2008(635024):8 pages, 2008.

[9] M. Åslund, E. Fredenberg, B. Cederström, and M. Danielsson. Spectral shaping for photon counting digital mammography. *Nucl. Instrum. Methods A*, 580(2):1046–1049, 2007.

[10] M Lundqvist, B Cederström, V Chmill, M Danielsson, and B Hasegawa. Evaluation of a photon-counting X-ray imaging system. *IEEE Trans. Nucl. Science*, 48(4):1530–1536, 2001.

[11] M. Lundqvist, B. Cederström, V. Chmill, M. Danielsson, and D. Nygren. Computer simulations and performance measurements on a silicon strip detector for edge-on imaging. *IEEE Trans. Nucl. Science*, 47(4):1487–1492, 2000.

[12] The HighReX Project. online: http://www.highrex.eu.

[13] J.M. Boone. Glandular breast dose for monoenergetic and high-energy x-ray beams: Monte carlo assessment. *Radiology*, 203:23–37, 1999.

[14] M.J. Berger, J.H. Hubbell, S.M. Seltzer, J.S., Coursey, and D.S. Zucker. XCOM: Photon Cross Section Database. online: http://physics.nist.gov/xcom. National Institute of Standards and Technology, Gaithersburg, MD, 2005.

[15] B.L. Henke, E.M. Gullikson, and J.C. Davis. X-ray interactions: photoabsorption, scattering, transmission, and reflection at E=50-30000 eV, Z=1-92. *Atomic Data and Nuclear Data Tables*, 54(2):181–342, 1993.

[16] J.M. Boone, T.R. Fewell, and R.J. Jennings. Molybdenum, rhodium, and tungsten anode spectral models using interpolating polynomials with application to mammography. *Med Phys*, 24(12):1863–74, 1997.